\documentclass[aps,prl,epsfig,twocolumn,showpacs,preprintnumbers,amsmath,amssymb,superscriptaddress,floatfix]{revtex4}
\usepackage{epsfig} \usepackage{dcolumn} \usepackage{bm}
\def\CuII{Cu$^{2+}$ }

\def\kagome{kagom\'{e} }

\begin{document}

\title{On the magnetic ground state
  of an experimental $S=1/2$ kagom\'{e} antiferromagnet.}
\author{M.~A.~de Vries} 
\affiliation{
  CSEC and School of Chemistry, The University of Edinburgh, Edinburgh,
  EH9 3JZ, UK}
\author{K.~V.~Kamenev} 
\affiliation{
  CSEC and School of Engineering \& Electronics, The University of Edinburgh,
  Edinburgh, EH9 3JZ, UK} 
\author{W.~A.~Kockelmann}  
\affiliation{ ISIS,
  STFC Rutherford Appleton Laboratory, Chilton, Didcot, OX11 0QK, UK}
\author{J.~Sanchez-Benitez}
\affiliation{
  CSEC and School of Engineering \& Electronics, The University of Edinburgh,
  Edinburgh, EH9 3JZ, UK} 
\author{A.~Harrison} \email{harrison@ill.eu}
\affiliation{ 
  Institut Laue-Langevin, 6 rue Jules
  Horowitz, F-38042 Grenoble, France} 
\affiliation{
  CSEC and School of Chemistry, The University of Edinburgh, Edinburgh,
  EH9 3JZ, UK} 

\date{\today}

\begin{abstract}
We have carried out neutron powder-diffraction measurements on zinc
paratacamite Zn$_x$Cu$_{4-x}$(OH)$_6$Cl$_2$ with $x=1$, and studied
the heat capacity in fields of up to 9~T for $0.5 \leq x \leq 1$.  The
$x=1$ phase has recently been shown to be an outstanding realisation
of the $S=1/2$ kagom\'{e} antiferromagnet. A weak mixing of
Cu$^{2+}$/Zn$^{2+}$ between the Cu and the Zn sites, corresponding to
$\sim 9$\% of all Cu$^{2+}$ for $x=1$, is observed using neutron
diffraction. This ``antisite disorder'' provides a consistent
explanation of the field dependence of the heat capacity for $0.8 \leq
x \leq 1$. From comparison of the derived Cu$^{2+}$ occupancy of the
Zn sites for $x = 0.8\ldots 1$ with the magnetic susceptibility, we
argue that for $x = 0.8\ldots 1$ zinc paratacamite is a spin liquid
without a spin gap. The presence of unpaired but nevertheless strongly
interacting spins gives rise to a macroscopically degenerate ground
state manifold, with increasingly glassy dynamics as $x$ is lowered.
\end{abstract}


\maketitle{}

Physical realisations of the $S=1/2$ kagom\'{e} Heisenberg
antiferromagnet have been long sought after because it is expected
that the ground state of this system can retain the full symmetry of
the underlying effective magnetic
Hamiltonian~\cite{Anderson:87,Lhuillier:2004}; the geometry of the
kagom\'{e} lattice frustrates the classical N\'{e}el antiferromagnetic
ordering, and no symmetry-breaking transition is expected even at
$T=0$~\cite{Leung:93,Mila:98,Waldtmann:98,Sind:00,Bernhard:02}. It has
been suggested that even in the thermodynamic limit the symmetric
quantum-mechanical electronic ground state is protected from
quantum-mechanical dissipation~\cite{Leggett:87} by a gap between the
non-magnetic ground state and the lowest magnetic (triplet)
excitations~\cite{Misguich:05b,Ioffe:02}. In practice, however, a gap
of order $J/20$~\cite{Waldtmann:98} may be too small to expect a
non-magnetic ground state in real materials.

\begin{figure}[ht]
\begin{picture}(245,196)(0,0)
\put(-7,88){\epsfig{file = 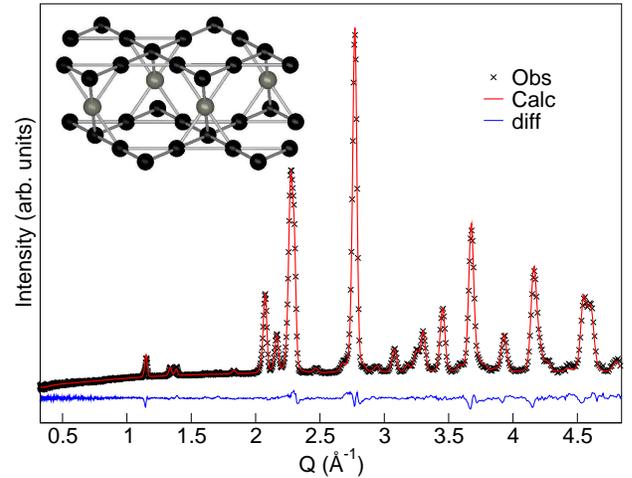, angle=0, width=2in}}
\put(0,0){\epsfig{file= Rotax10Kani.eps ,angle=0,width=3.2in}}
\end{picture}
\caption{Section of the neutron diffraction pattern of
ZnCu$_3$(OD)$_6$Cl$_2$ at 10~K, and the result of the structure
refinement (residuals $\chi ^2 = 13.95$, $R_p = 2.83\%$, $R_{wp} =
2.73\%$).  The inset shows two kagom\'e layers with Cu sites (black),
and the interplane Zn site (grey) of the zinc paratacamite
structure. The antisite disorder means that a fraction of the
Cu$^{2+}$ ions interchange with (for $x=1$ an equal number of)
Zn$^{2+}$ ions.}
\label{figure:rotax10K}
\end{figure}

Shores~\emph{et al.}~\cite{Shores:05} have shown that in the Cu salt
herbertsmithite~\cite{Braithw:04} (ZnCu$_3$(OH)$_6$Cl$_2$, depicted in
the inset of figure~\ref{figure:rotax10K}) antiferromagnetically
coupled Cu$^{2+}$ ions are located at the vertices of a kagom\'{e}
lattice. Muon experiments have shown that the ground state of this
system is either paramagnetic or (quantum) spin-liquid. Almost no muon
relaxation was observed even at 50~mK~\cite{Mendels:07}, despite the
large Weiss temperature $\theta_{\rm{w}}\approx
-300$~K~\cite{Shores:05,Helton:07}. Separating the kagom\'{e} layers
are Zn sites of O$_{\rm{h}}$ symmetry, which can also host Cu$^{2+}$
ions to form the zinc paratacamite family of stoichiometry
Zn$_x$Cu$_{4-x}$(OH)$_6$Cl$_2$ with $0.3 \leq x \leq 1$. For Zn$^{2+}$
stoichiometries $x<0.3$, the Zn site is mainly occupied by Jahn-Teller
active Cu$^{2+}$ ions and becomes angle distorted. At this point the
symmetry of the lattice changes from rhombohedral ($x>0.3$) to
monoclinic, forming the end-member clinoatacamite~\cite{Zheng:05}. 
Due to the strong Jahn-Teller distortion of the Cu sites on the kagom\'{e}
lattice, the mixing of Cu$^{2+}$ and Zn$^{2+}$ between the Cu and the Zn
sites (antisite disorder) in the $x=1$ phase can be expected to be
low, but exactly how low has not previously been measured. From the
magnetic susceptibility of samples with $x<1$ it is clear that the
Cu$^{2+}$ ions on the inter-plane Zn site are only weakly coupled to
the kagom\'{e} layers, and it has been suggested that the divergence
of the magnetic susceptibility for $x=1$ at low temperatures can be
explained by antisite permutations of 6-7\% of the Cu$^{2+}$ ions with
$\sim 19$\% of the Zn$^{2+}$ \cite{Ran:07,Misguich:07,Bert:07}.

To be able to account for the antisite disorder in the further
analysis of the system, the Cu$^{2+}$/Zn$^{2+}$ mixing was
measured using neutron powder diffraction at the Rotax neutron
time-of-flight diffractometer at the ISIS facility, United Kingdom. We
used a 4~g powder sample of deuterated $x=1$ zinc paratacamite,
synthesised using the hydrothermal method as described
in~\cite{Shores:05}. The purity of the samples and Cu$^{2+}$ to
Zn$^{2+}$ ratio were verified with powder x-ray diffraction and
Inductively Coupled Plasma Auger Electron Spectroscopy (ICP-AES) with
an accuracy of $\pm0.03$ in $x$. Neutron-diffraction data were
collected at 285, 150 and 10~K, and Rietveld analysed against the
structure as reported in~\cite{Shores:05}. No structural changes were
observed with temperature, and the level of deuteration was refined to
94.0(6)\%. During the refinement the total Cu$^{2+}$:Zn$^{2+}$ ratio
was fixed at 3:1, as measured directly using ICP-AES. It was further
assumed there were no vacant Zn or Cu sites. In this way the Cu$^{2+}$
occupancy on the kagom\'{e} lattice was refined to $91(2)\%$,
corresponding to a Zn$^{2+}$ occupancy of the inter-plane Zn site of
$73(6)\%$~cite{footnote1}, for the highest-statistics data set taken at
10~K. The result is given in figure~\ref{figure:rotax10K}. The
refinements of the data taken at 150 and 285~K were in overall
agreement within the experimental error. On relaxation of the
constraints the solution was stable, but no longer unique. There was a
small further reduction of the residues for a slight reduction of the
Cu and Zn site occupancies, which also suggests that the experimental
error will be larger than stated.

\begin{figure}[htbp]
\begin{center}
  \epsfig{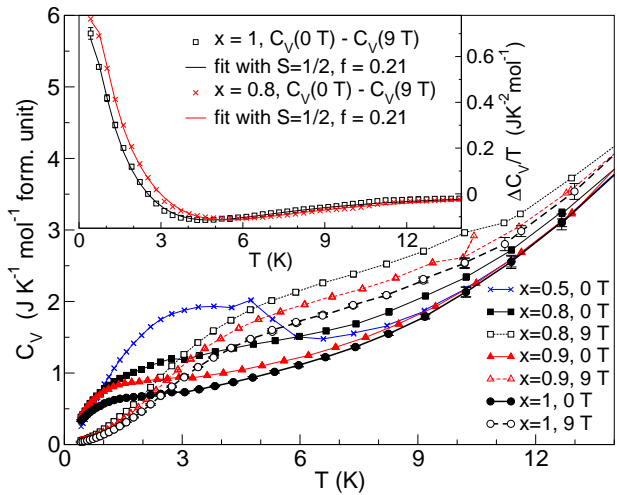}
\caption{\small{The heat capacity in 0~T for samples with $x=0.5$,
    0.8, 0.9 and 1 as well as in 9~T for $x=0.8$, 0.9 and 1. The error bars
    are given for $x=1$ only. The inset displays $\Delta C_V/T$ for
    $x=0.8$ and $x=1$ and their respective fits. }}  
\label{figure:Znxfits}
\end{center}
\end{figure}

\begin{figure}[htbp]
\begin{center}
\epsfig{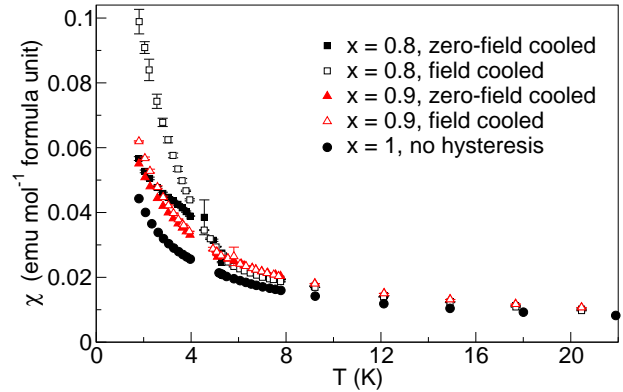}
\caption{\small{The zero-field cooled and field cooled magnetic
    susceptibility for zinc paratacamite with $x = 0.8$ and $x=0.9$,
    compared with the susceptibility for $x=1$ which has a negligible
    hysteresis. }}
\label{figure:hyst}
\end{center}
\end{figure}

The heat capacity measurements were carried out using a Quantum Design
PPMS system, on $\sim 5$~mg dye-pressed pellets of
Zn$_x$Cu$_{4-x}$(OH)$_6$Cl$_2$ with $x = 0.5, 0.8, 0.9$
and 1.0. We could reproduce the heat capacity for $x=1$ in 0, 1, 2, 3,
5, 7 and 9~T fields as reported by
Helton~\emph{et~al.}~\cite{Helton:07}. Figure~\ref{figure:Znxfits}
presents the heat capacities of samples with $x=0.8,0.9$
and $x=1$ in 0 and 9~T and for $x=0.5$ in 0~T. For intermediate
fields, not shown here for clarity, the shoulder gradually moves to
higher temperatures while the total entropy below $\sim 24$~K remains
constant. The magnetic susceptibility of samples with $x=0.8$, $x=0.9$
and $x=1$ as shown in figure~\ref{figure:hyst} was measured with a
Quantum Design MPMS system, on $\sim 50$~mg pellets.

\begin{figure}[ht]
\begin{center}
\epsfig{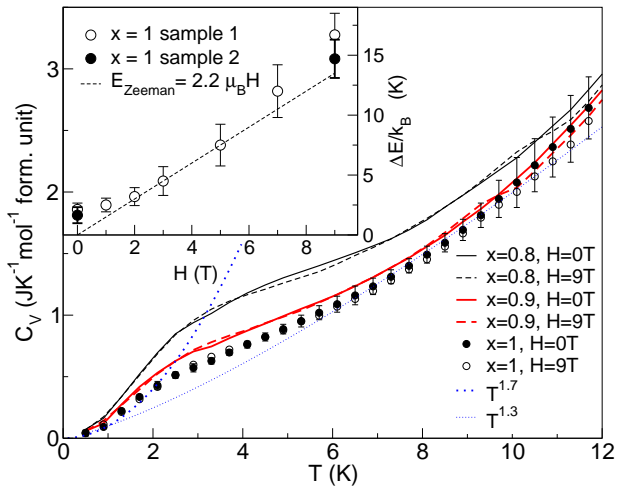}
\caption{\small{The field-independent part of the heat capacity as
obtained from the 0~T and 9~T data of samples with $x = 0.8$, 0.9 and
1. The dotted lines (blue online) give fits to the corrected heat capacity for $x=1$
with $\gamma T^{\alpha}$ with $\alpha =1.3$ and 1.7. The inset shows
$\Delta E$ as a function of $H$ between 0 and 9~T, compared to the
Zeeman splitting with $g=2.2$. Sample 2 is the deuterated $x=1$ sample
which was used in the neutron diffraction experiment.}} 
\label{figure:CvremZnx}
\end{center}
\end{figure}

To the eye, the field dependence of the heat capacity is similar to
the Schottky anomaly arising from defects in Zn-doped Y$_2$BaNiO$_5$
and [Ni(C$_2$H$_8$N$_2$)$_2$(NO$_2$)]ClO$_4$
(NENP)~\cite{Ramirez:94}. Hence, we have applied a similar analysis as
described in~\cite{Ramirez:94}.  To study the field-dependent part of
the heat capacity, for each sample ($x$) the difference was taken
between the interpolated heat-capacity curves measured in different
fields. The inset in figure~\ref{figure:Znxfits} shows the difference
between the 0 and 9~T curves $\Delta C_V/T = [C_V(H_1=0~$T$) -
C_V(H_2=9~$T)$]/T$ for $x=0.8$ (crosses) and for $x=1$ (squares). We
found that the field-dependent part of the heat capacity can be
modelled by a small number of zero-field split doublets,
i.e. interacting $S=1/2$ spins or $S=1/2$ excitations. $\Delta C_V/T$
was fitted with $f [ C_V^{S=1/2}(\Delta E_{H1}) - C_V^{S=1/2}(\Delta
E_{H2})]/T$, where $f$ is the fraction of doublets per unit cell (or
their spectral weight). $C_V^{S=1/2}(\Delta E_H)$ is the heat capacity
from a $S=1/2$ spin with a level splitting
$\Delta E_H$ which for fields $H \geq 2$~T equals the Zeeman splitting
with $g \approx 2.2$, as shown in the inset of
figure~\ref{figure:CvremZnx}. The shoulder in the heat capacity in
zero-field, which corresponds to a zero-field splitting of the
doublets of $\Delta E\sim 1.7$~K (0.15~meV) for $x=1$,  $\Delta E\sim
2.1$~K for $x=0.9$ and $\Delta E\sim 2.2$~K for $x=0.8$ indicates that
the levels involved are part of an interacting system, and cannot be
ascribed to a paramagnetic impurity phase. The best agreement with
experiment was obtained when a small Gaussian spread $\sigma$ in level
splittings $\Delta E$ was taken into account, indicated as the error
bars in the inset of figure~\ref{figure:CvremZnx}. This brings the
number of fit parameters to 5. The lines through
the data points in the inset of figure~\ref{figure:Znxfits} are the
fit results for $x=0.8$ and $x=1$. We find that
$f=0.21(1), 0.22(1)$ and 0.19(1) for $x=0.8, 0.9$ and $1.0$
respectively. For $x=1$, with three Cu$^{2+}$ ions per unit cell, this
accounts on average for 6.0(6)\% of all Cu$^{2+}$.

Due to a spin gap the heat capacity of the $S=1/2$ kagom\'e
antiferromagnet is expected to show a shoulder between $J/20$ and
$\sim J/10$ corresponding to the population of the lowest magnetic
($S_{\rm{tot}}=1$) levels~\cite{Waldtmann:98, Sind:00, Bernhard:02,
Misguich:05}. In our data a shoulder is evident in zero
field. However, with the application of a magnetic field this shoulder moves to
higher temperatures (energies), as shown in the inset of
figure~\ref{figure:CvremZnx}. This is very different from what can be
expected for, for example, a singlet-triplet system with a non-magnetic ground
state. For the latter an applied field will lower the energy of the
$S=1$ level with spin aligned along the field, so that for
sufficiently strong fields a level crossing occurs and this $S=1$
level will become the ground state. It is clear that such a level
crossing is not observed here. Most likely, the lowest energy level involved in
the system giving rise to a field-dependence is a magnetic level
too. Several models have been tried, of which only the doublet (a
$S=1/2$ system) gives an overall consistent fit for all 18 curves from
a total of 5 samples, for each model using only 5 fitting
parameters. A model with a triplet of $S=1$ levels results in a
slightly poorer fit to the data as compared with a doublet. Similar
models with higher-level multiplets ($S_{\text{tot}} > 1$) can not be
brought into agreement with our data. It should be noted that a
doublet with a field-dependence as described here has also been
observed in neutron spectroscopy data~\cite{Helton:07,SHLee:07}. As is shown in
figure~\ref{figure:hyst}, the system gradually develops a magnetic
hysteresis as $x$ is lowered (the Cu$^{2+}$ concentration is
increased), while the muon relaxation increase is indicative of a
slowing-down of the spin dynamics~\cite{Mendels:07}. The hysteresis is
a history dependence which rules out a macroscopic quantum state for
the system as a whole, since such a state would have a unitary time
evolution as described by the Schr\"odinger equation. That the latter is not
the case here is also clear from the energy gap for the antisite spins
in zero field, which increases as $x$ is lowered. Using our model this
increase is quantified as a gap of $1.7$~K for $x=1$ to $2.2$~K for
$x=0.8$. This may be the strongest indication that the energy gap
corresponds to local excitations rather than coherent many-body
quantum states of the total system. In the latter case the time
dependence should follow the Schr\"odinger equation where a larger gap
leads to faster dynamics. 

We suggest, as is also done in~\cite{Ran:07,Misguich:07,Bert:07}, that
the fraction $f$ of zero-field split doublets, which models the
field-dependence in the heat capacity for $0.8 \leq x \leq 1$, are
weakly-coupled $S=1/2$ spins from Cu$^{2+}$ ions residing on
inter-plane Zn sites (antisite spins). For $x=1$ an identical fraction
$f$ of Zn$^{2+}$ ions must occupy Cu sites on the kagom\'{e}
lattice. Once $f(x)$ is known the Cu$^{2+}$ coverage $c(x)$ of the
three Cu sites per unit cell is given by $c=4-x-f(x)$. An important
assumption in our argument is that the heat capacity of a slightly
diamagnetically doped kagom\'{e} lattice is field independent, which
is reasonable as long as $g \mu_{\rm{B}} H \ll
\theta_{\rm{w}}$~\cite{Ram:00Cv_SCGO,Sind:00}. For the deuterated
$x=1$ sample of the neutron-diffraction measurements it follows that
the antisite disorder is 6.3(3)\% in Cu$^{2+}$ or 19.0(9)\% in
Zn$^{2+}$, in rough agreement with the neutron diffraction result.

\begin{table}[htbp]
\caption{\small{The fitted fraction of antisite spins per
unit cell $f$, the corresponding Cu$^{2+}$ occupancy of the kagom\'{e} lattice
$c=4-x-f$, the total entropy from the antisite spins $\mathcal{S}_f$,
the measured total entropy $\mathcal{S}(T)$ up to 24~K and the
percentage of the entropy recovered per Cu$^{2+}$ spin.}}
\begin{center}
\begin{tabular}{l l l r r r } 
\hline \hline
$x$ & $f$ & $c$ &$\mathcal{S}_f~/R$ & $\mathcal{S}(T)~/R$ & 
$\frac{\mathcal{S}(T)}{(4-x)\ln(2)}$ 
\\
\footnotesize{$\pm 0.03$}&\footnotesize{$\pm 0.018$} 
&\footnotesize{$\pm 0.03$} & \footnotesize{$f\ln(2)$} &
\footnotesize{@$T = 24$~K~$^b$ } & \footnotesize{/\%}\\
\hline
0.50 & 0.50$^a$& 3.00 & 0.346(8) & 1.061(12) & 43.9(2)\\
0.80 & 0.210 & 2.97 & 0.15 & 0.993(11) & 44.7(2)\\
0.90 & 0.220 & 2.88 & 0.15 & 0.959(9)  & 44.8(2)\\
1.00 & 0.190 & 2.81 & 0.13 & 0.933(9)  & 44.8(2)\\
\hline\hline
\end{tabular}
\end{center}
\footnotesize{
$^a$~Here $f$ was not obtained from the heat capacity which at this
  level of doping is altered due to a cooperative transition. However,
  for $x=0.5$ it is safe to assume that $c=3.0$ (full occupancy) and
  hence $f = x$. $^b$~For $T>24$~K no relative changes in
  $\mathcal{S}(T)$ occur between samples with different $x$. 
}
\label{table:Stot}
\end{table}

Comparing the heat-capacity data of several $x=1$ samples, all
synthesised at a temperature of 484~K, an average antisite disorder of
$\sim 6.0(6)$\% in Cu$^{2+}$ is derived, as listed in
table~\ref{table:Stot}, along with the results for $x=0.8$ and
$x=0.9$. The chemical potential behind the Cu$^{2+}$/Zn$^{2+}$
partitioning can now be estimated to $\sim 1400$~K, a plausible value
given that most likely the Zn site becomes locally slightly
angle-distorted, if occupied by an otherwise orbitally degenerate
Cu$^{2+}$ ion. 
The Cu sites on the
kagom\'{e} lattice are energetically favoured by the Cu$^{2+}$ ions,
and there is only a slow increase of the Cu$^{2+}$ occupancy on the Zn
sites until the Cu$^{2+}$ occupancy of the kagom\'{e} lattice ($c$ in
table~\ref{table:Stot}) is almost complete. For $x=0.8$ the magnetic
hysteresis is too large to be ascribed to impurities or local
variations in Zn stoichiometry (Fig.~\ref{figure:hyst}). Since even at
$x=0.8$ only $\sim20$\% of the Zn sites are occupied by Cu$^{2+}$, this
hysteresis must be due to the higher connectivity of the 3D lattice,
which is mainly due to the higher \CuII occupancy of the kagom\'e
planes (see table~\ref{table:Stot}). Hence, for the phases $x<1$ which
have a magnetic hysteresis the kagom\'e layers must be in a magnetic
state, i.e. both singlet and triplet states mix into the ground
state. Since no quantum phase transition occurs between $x=0.8$ and
$x=1$ as is clear from our heat capacity data, the ground state of the
kagom\'e layers in the $x=1$ phase must be magnetic too. This is in
support of NMR measurements~\cite{Imai:07,Olariu:08} in that there is
no spin gap. What is remarkable in the present case is that the
appearance of unpaired spins precedes the breaking of spin-rotational
symmetry to a long-range ordered state. This results in a
macroscopically degenerate ground state with increasingly glassy
dynamics as $x$ is lowered.  

The heat capacity of the kagom\'{e} lattice can be estimated by
subtracting the heat capacity from the Cu$^{2+}$ spins on the Zn
sites. The result for the data with $0.8\leq x \leq 1$ is shown in
figure~\ref{figure:CvremZnx}. For all $x$ the curves obtained from the
0 and 9~T data are identical within the experimental error, which
follows from the quality of the fit as described in the previous
paragraphs. This part of the heat capacity most likely corresponds to
the kagom\'{e} layers. In this field-independent part of the heat
capacity a weak shoulder is visible at a slightly higher temperature
than the shoulder due to the antisite spins. As $x \to 1$ the shoulder becomes less
pronounced, and hence, it may be interpreted as due to the entropy
release when the fluctuations in neighbouring \kagome layers, which
are connected via the antisite spins, decouple. If the heat capacity
from perfectly 2D \kagome layers follows a power-law for $T \to 0$,
then taking into account the distortive effect of the shoulder due to
couplings between the \kagome layers, $0.1T^{\alpha}$~mol$^{-1}$
formula units with $\alpha=1.3(1)$ is our best estimate. An exponent
$\alpha = 2$ cannot be brought into agreement with our data.

In summary, based on the field-dependence of the shoulder in the
low-temperature heat capacity, which corresponds to the weakly
dispersive feature observed at the Zeeman energy in neutron
spectroscopy data, we rule out interpretations based on a
singlet-triplet splitting. Rather, the feature remains a doublet over
the entire range of applied fields. We suggest these doublets are the
magnetic states of the \CuII antisite spins, which raises an important
question as to the origin of the observed zero-field splitting of the
antisite spins. From analysis of the heat capacity and the magnetic
susceptibility as a function of $x$ we further conclude that even for
$x=1$ the ground state of the kagom\'e system is a gapless spin
liquid. 

We gratefully acknowledge helpful discussions with Philippe Mendels
(Universit\'e Paris Sud), Paul Attfield and Philippe Monthoux (the
University of Edinburgh), Claudine Lacroix (Lab. Louis N\'eel),
Andreas L\"auchli and Henrik R\o{}nnow (EPFL). We further
acknowledge an EPSRC grant - EP/E06471X/1. MdV also thanks the ESF HFM
network for an exchange grant.


\begin{thebibliography}{10}
\bibitem{footnote1}
The stated error bar is $3\times$ the $\sigma$
estimated in the Rietveld refinement, corresponding to a 99\%
confidence interval.
\end{thebibliography}
\end{document}